\documentstyle[12pt,amssym,aasms4,epsf]{article}
\received{September 7, 2000}
\accepted{December 5, 2000}

\slugcomment{}

\lefthead{Kulkarni et al.}
\righthead{NICMOS Imaging of a Damped Ly-$\alpha$ Absorber at $z=1.86$}

\begin{document}

\title{A Search for the Damped Ly-$\alpha$ Absorber at $z=1.86$ \\
toward QSO 1244+3443 with NICMOS}

\author{Varsha P. Kulkarni\altaffilmark{1}, John M. Hill, Glenn Schneider}
\affil{University of Arizona, Steward Observatory, Tucson, AZ 85721}
\author{ Ray J. Weymann, Lisa J. Storrie-Lombardi\altaffilmark{2},}
 \affil{Carnegie Observatories, Pasadena, CA 91101}
\author{Marcia J. Rieke, Rodger I. Thompson,} 
\affil{University of Arizona, Steward Observatory, Tucson, AZ 85721}
\and 
\author{Buell T. Jannuzi}
\affil{National Optical Astronomy Observatories, P. O. Box 26732, Tucson, AZ 
85726-6732 }
\altaffiltext{1} {Present address: University of South Carolina, Dept. of 
Physics \& Astronomy, Columbia, SC 29208}
\altaffiltext{2} {Present address: SIRTF Science Center, Caltech, Pasadena, CA 91125}

\vskip 1.0in
\centerline{Accepted for Publication in The Astrophysical Journal}


\begin{abstract}
We have carried out a high-resolution imaging search for the galaxy 
associated with the damped Lyman-$\alpha$ (DLA) absorber at $z=1.859$ 
toward the $z_{em}=2.48$ quasar QSO 1244+3443, using the Hubble Space 
Telescope (HST) and the Near Infrared Camera and Multiobject 
Spectrometer (NICMOS).  
Images were obtained in the broad filter F160W and the 
narrow filter F187N with camera 2 on NICMOS with the goal of detecting 
the rest-frame optical continuum and the H-$\alpha$ line 
emission from the DLA. After PSF subtraction, two weak features 
are seen at projected separations 
of 0.16-0.24" from the quasar. Parts of these features may be 
associated with the 
DLA absorber, although we cannot completely rule out that they could be  
artifacts of the point spread function (PSF). If associated with the DLA, the objects would be 
$\approx 1-2  \,  h_{70}^{-1}$ kpc in size with integrated flux densities 
of 2.5 and 3.3  
$\mu$Jy in the F160W filter, implying luminosities at 
$\lambda_{central}=5600$ {\AA} in the DLA rest frame of 
$4.4-5.9 \times 10^{9} h_{70}^{-2}$ $L_{\odot}$ at $z=1.86$, for 
$q_{0}=0.5$. However, no significant H-$\alpha$ line emission is seen from 
these objects, suggesting low star formation rates (SFRs). Our 3 $\sigma$ 
upper limit on the SFR in the DLA is 1.3 h$_{70}^{-2}$ M$_{\odot}$ yr$^{-1}$ 
for q$_{0} = 0.5$ (2.4 h$_{70}^{-2}$ M$_{\odot}$ yr$^{-1}$ for q$_{0} = 0.1$). 
This together with our earlier result for LBQS1210+1731 mark a significant 
improvement over previous constraints on the star formation rates of DLAs. 
Dust within the DLA could extinguish H-$\alpha$ emission, but  
this would require the dust content in the DLA to be much higher than 
that inferred from previous DLA observations. A combination of low 
star formation rate and some dust extinction is likely to be responsible 
for the lack of H-$\alpha$ emission. Alternatively, the objects, 
if real, may be associated 
with the host galaxy of the quasar rather than with 
the DLA absorber. In any case, our observations suggest 
that the DLA is not a large bright proto-disk, but a compact object or 
a low-surface brightness galaxy. 
If the two features are PSF artifacts or associated with the quasar 
host, then the constraints on the size and luminosity   
of the DLA are even more severe. 

\end{abstract}


%

\newpage
\section{INTRODUCTION}

Damped Ly-$\alpha$ absorption (DLA) systems in quasar spectra contribute 
significantly to the neutral hydrogen content of the universe and are thought 
to be the progenitors of present-day galaxies. However the exact nature 
of DLAs is not yet understood. DLAs are variously thought to be 
proto-spirals (Wolfe et al. 1986; Prochaska \& Wolfe 1997, 1998), 
gas-rich dwarf galaxies (York et al. 1986; Matteucci et al. 1997), 
merging proto-galactic fragments in cold dark matter cosmologies  
(e.g., Haehnelt et al. 1998), collapsing halos with merging clouds  
(e.g., McDonald \& Miralda-Escud\'e 1999), or low-surface brightness galaxies 
(Jimenez, Bowen, \& Matteucci 1999). It has not been possible to 
determine which of these scenarios hold(s) true, because of the 
difficulty in detecting the emission from high-redshift DLAs. 

Most attempts to detect the Ly-$\alpha$ emission from high-redshift 
intervening ($z_{abs} < z_{em}$) DLAs have produced either 
non-detections or weak detections (e.g. Smith et al. 1989; Hunstead, 
Pettini, \& Fletcher 1990; Lowenthal et al. 1995; Djorgovski et al. 1996; 
Djorgovski 1997). However these data  
do not effectively constrain the star formation rates (SFRs) in the DLAs, 
since even small quantities of dust can extinguish the Ly-$\alpha$ 
emission, owing to the resonant scattering of the Ly-$\alpha$ 
photons (Charlot \& Fall 1991). Small but significant amounts of 
dust are indeed inferred to exist in DLAs from observations 
of reddening of background quasars and heavy element depletion 
(e.g., Pei, Fall, \& Bechtold 1991; Pettini et al. 1997; Kulkarni, 
Fall, \& Truran 1997). 
There have been attempts to detect DLAs in longer wavelengths such 
as H-$\alpha$ which are less affected by dust and not subject to 
resonant scattering. However, these have either resulted in non-detections 
or have only detected companions separated by large angular distances from 
the quasars (see, e.g, Teplitz, Malkan, \& McLean 1998; 
Bechtold et al. 1998; Mannucci et al. 1998, Bunker et al. 1999). 
The detection of H-$\alpha$ emitters close to the quasar 
sightline has not been possible in high-$z$ intervening DLA fields. 
A summary of the previous attempts to detect high-$z$ DLAs in emission 
is given by Kulkarni et al. (2000).

With the goal of searching for H-$\alpha$ emitting objects at close 
separations from the quasar, we have obtained deep diffraction-limited  
images of three DLAs at $z \sim 2$ using the Near Infrared Camera and 
Multi-Object Spectrometer (NICMOS) onboard the Hubble Space Telescope 
(HST). In an earlier paper, we described the observations of the 
$z=1.89$ DLA toward LBQS 1210+1731 in detail (Kulkarni et al. 2000). 
Here we describe our observations of the quasar 1244+3443 
($z_{em}=2.48$), which has a 
spectroscopically known damped Ly-$\alpha$ absorber ($z_{abs}=1.859$ 
and log $N_{HI} = 20.6$; Wolfe et al. 1995). Our observations combine 
high near-IR sensitivity and high spatial 
resolution with a more stable PSF than is currently 
possible with ground-based observations. Some of 
our observations also use the NICMOS coronagraph, which greatly 
decreases the scattered light background 
outside of the coronagraphic hole. Sections 2, 3 and 4 describe the 
observations, data reduction, and the subtraction of the quasar point 
spread functions. Our results are described in section 5. 
Finally, sections 6 and 7 summarize the constraints 
from our observations on sizes, environment, and SFRs of DLAs.

\section{OBSERVATIONS}
The field of QSO 1244+3443 was observed at two epochs (1998 
August 2 04:31-12:29 UT and 1998 July 5 18:32-21:22 UT), 
using NICMOS camera 2 (image scale $\sim 0.07598 \arcsec \times 0.07530 
\arcsec$, field 
of view $19.53 \arcsec \times 19.3 \arcsec$). 
A sequence of spatially offset broad-band images was obtained in 
MULTIACCUM mode with the F160W (H) filter (central wavelength 
1.5940 $\mu$m, FWHM 0.4030 $\mu$m). Field offsetting was accomplished 
with a 4-point spiral dither pattern in steps of $\approx 7.5$ pixels, 
using the Field Offset Mirror (FOM) internal to NICMOS. The exposures 
at each dwell point were 640 s long, giving a total integration time 
of 2560 s. The MULTIACCUM observations 
consisted of non-destructive readouts 
in the ``STEP64'' readout timing sequence, i.e. ``MULTIACCUM'' readouts 
separated logarithmically up to 64 s and linearly in steps of 64 s beyond 
that. In addition, narrow-band images were obtained in the filter F187N 
(central wavelength 1.875 $\mu$m, FWHM 0.0188 $\mu$m), 
in which the redshifted H-$\alpha$ emission from the DLA, if present,  
would lie.  Four-point spiral dither patterns in steps of 7.5 pixels, 
with a 704 s STEP64 MULTIACCUM exposure at each dwell point were 
repeated in four successive orbits, resulting in a total integration 
time of 11,264 s.  The spatial resolution 
of the F160W and F187N images is 0.14 $\arcsec$ (1.8 pixels) and 0.17 
$\arcsec$ (2.1 pixels) FWHM, respectively. Thus, critical sampling is 
achieved with both filters by combining the dithered images onto a nearly-
optimal (half pixel sub-sampled) grid. 

Broad-band images in the F160W filter were also obtained 
using the camera 2 coronagraph on 1998 July 5 from 18:32 to 21:22 UT. 
These consisted of an initial pair of 160 s long target-acquisition 
images, which were followed by 
placement of the quasar in the coronagraphic hole (0.3 $\arcsec$  or 4 
pixels in geometrical radius) and then integration for a 
total of 5184 s (3 exposures of 768 s each in the first orbit and 3 
exposures of 960 s each in the second orbit, all using the STEP64 
MULTIACCUM timing sequence). No dithering was used, of course, for 
the coronagraphic observations. The coronagraphic system 
significantly reduces both scattered and diffracted energy from the 
occulted target's point spread function core by factors of 4-6 in 
the region of $0.55 \arcsec < r < 1.5 \arcsec$, 
compared to direct imaging (Schneider et al. 1998; 
Lowrance et al. 1998). 

To circumvent image artifacts known as ``bars'' in all our camera 2 
images, cameras 1 and 3 were run in parallel.

\section{REDUCTION OF IMAGES}

The images were reduced using the IRAF package NICRED 1.8 
for the reduction of MULTIACCUM NICMOS data (McLeod 
1997). The dark images used were made from 
on-orbit dark exposures taken during the NICMOS calibration program. 
For the noncoronagraphic images, the flat-field image used was 
made from on-orbit exposures taken with the internal calibration 
lamps during the NICMOS Cycle-7 calibration program. For the coronagraphic 
images, the flat-field image was made with target-acquisition data 
taken just before the coronagraphic exposures. This flat is far 
better, for regions close to 
the edge of the coronagraphic hole (and even out to $\approx 1 \arcsec$), 
than the standard reference flats, which are severely affected by 
hole-edge gradients due to the relative motion of the hole. Use of 
the flat made with the target acquisition data ensures that 
the coronagraphic hole is in the same position on the detector as the 
quasar data, which is critical for studying faint objects close 
to the edge of the coronagraphic hole. See Kulkarni et al. (2000) for 
further details on the sequence of data reduction tasks performed by 
NICRED 1.8.

The images for the different dither positions 
were registered by cross-correlating with the IRAF task XREGISTER. 
The quasar was used as the reference object since it was the only point 
source available in our images. 
Finally, the registered images were averaged together using a bad-pixel 
mask that took out any remaining bad pixels, and rejecting pixels 
deviating by more than 3 $\sigma$ from the average of the five F160W 
images, using averaged sigma-clipping.

For the F187N images, where there were four exposures (one in each orbit) 
at each of the four dither positions, we first median-combined the four  
exposures at each position separately, and then registered and 
averaged the four positions together to make the final image.
For the coronagraphic F160W images, where there were four exposures at 
the same position in each of the two orbits, we averaged the 
exposures in each orbit separately and then took a weighted average 
(weighting by exposure times) of the combined exposures from the 
two orbits. 

Figs. 1, 2, and 3 show the final reduced images for the 
noncoronagraphic F160W, noncoronagraphic F187N, and coronagraphic 
F160W observations. The orientations of Figs. 1 and 2 agree 
exactly while they differ from that of Fig. 3 by 19.046 degrees. 
The color table and the intensity scale in ADU s$^{-1}$ are shown 
below each figure.  Figures 1 
and 2  show the quasar point source along with the diffraction pattern. 
The quasar light has been reduced greatly (although not completely) in the coronagraphic image in Fig. 3. A number of field galaxies are seen 
in the coronagraphic image at large angular separations from the 
quasar. To study whether there is any galaxy close to the quasar 
that may cause the DLA, we need to subtract the respective PSFs. 

\section{SUBTRACTION OF THE QUASAR POINT SPREAD FUNCTION}

To obtain reference point spread functions for subtraction, we used  
observations of stars in the same filter / aperture combinations as those 
employed for the quasar imaging. We did not include PSF star observations 
in our own observations since we wanted to maximize the use of the 
available HST observing time for imaging of the quasar fields. We 
therefore used PSF star observations from other programs (in particular 
the stellar images from the photometric monitoring program carried out 
during Cycle 7 NICMOS calibration) for constructing the reference 
PSFs for subtraction. 
 
For the noncoronagraphic images, we chose the PSF observations 
such that the telescope focus ``breathing'' 
(Bely 1993) values matched as closely as possible the 
values for the DLA observations. This attempts to minimize changes in 
the fine structure of the PSF caused by changes in the HST focus. For the 
noncoronagraphic F160W and F187N images, we used the PSF star P330E, 
observed on 1998 July 8 and 1997 August 5, respectively. The F160W 
noncoronagraphic image of P330E, made by 
combining four exposures of 3 s each, had a count rate of 108.10 
ADU s$^{-1}$ at the maximum of the first Airy ring. The corresponding 
quasar image, made by combining four exposures of 640 s each, 
had a count rate of 0.99 ADU s$^{-1}$ at the maximum of the first 
Airy ring. For the F187N filter, the P330E image, made by combining 
three exposures of 80 s each, had 3.76 ADU s$^{-1}$ at the maximum of 
the first Airy ring. The corresponding count rate was 0.037 
ADU s$^{-1}$ for the F187N quasar image, made by combining 
16 exposures of 704 s each.  See 
Kulkarni et al. (2000) for details on the selection of PSF 
stars with appropriate breathing values and for an in-depth discussion of 
the effect of breathing, color of PSF star, and other PSF details 
on the results of PSF subtraction. 

For the coronagraphic observations, the PSF 
star was chosen such that the position of the 
star in the coronagraphic hole be as close as possible to 
that of the quasar in our observations. This is important, because 
the PSF wings and ``glints'' 
from the edge of the coronagraphic hole depend sensitively 
on the precise position of the point source within the hole. 
We therefore used the observations of star GL577 for which 
we had coronagraphic observations (from another NICMOS GTO program), 
with the star placed at a position within 0.5 
pixels of the position of the quasar QSO 1244+3443 in our 
data. (This was the closest available observed coronagraphic PSF 
to our quasar observations. Unfortunately, 
PSF stars with better matching locations in the hole were not available.) 
The observations of GL577 were taken on 1998 
June 3 at a breathing value close to that for our 
quasar coronagraphic observations. 

The PSF star observations were analyzed 
in exactly the same manner as the quasar observations. The same 
interpolation scheme 
was used for resampling of the PSF star and quasar images. 
(We note that differences in the sampling of the 
quasar and PSF star images could potentially give rise to some 
artifacts. However, the overall conclusions of our study are 
unaffected by whether or not the quasar and PSF images are resampled. 
See Kulkarni et al. 2000 for further details.) The reduced PSF star 
images were subtracted from the corresponding 
quasar images after suitable scaling and registration,  
using the IDL program ``IDP-3'' (Lytle et al. 1999). The relative 
x and y alignment of the PSF star image with respect 
to the quasar image and the intensity scaling factor 
for the PSF star image were fine-tuned iteratively 
to obtain the minimum variance in roughly 
3 $\arcsec$  x 3 $\arcsec$ subregions around the quasar 
in the PSF-subtracted image. Radial flux plots of the quasar image, 
the aligned and scaled PSF image, and the difference of the two 
were also examined to check the alignment and scaling of the 
PSF. Figs. 4a, 5a, 6a show zoomed  
$\approx 2.7$\arcsec$  \times 2.7$\arcsec$ $ subregions around 
the quasar, from the noncoronagraphic F160W, noncoronagraphic 
F187N, and coronagraphic F160W images shown in Figs. 1, 2, 3, 
respectively. Figs. 4b, 5b, 6b show the PSF-subtracted versions 
of Figs. 4a, 5a, and 6a, respectively, using the closest 
matching PSFs available. 

\section{RESULTS}

\subsection{NONCORONAGRAPHIC F160W IMAGES}

Fig. 4b shows the F160W image after subtraction of the PSF image of 
star P330E dated 1998 July 8. The 
diffraction pattern disappears completely and most of 
the residual image contains a random mixture of positive and negative 
values. Two weak residuals are seen to remain near the quasar:  
One of the features is ``below'' the center, 
about 2 pixels (0.16 $\arcsec$) away from the center, while the 
second feature is to the ``lower right'' of the quasar, at a distance of 
about 0.24 $\arcsec$ from the quasar center. We name these features 
O1 and O2 respectively. (These features 
are seen more clearly if the data are sub-sampled by a factor of 2.) 
These features can not be made to disappear after reregistration of the 
PSF and quasar images or rescaling of the PSF image without causing large 
negative residuals elsewhere. 
We cannot completely rule out that O1 and O2 are 
artifacts in the PSF. However, given the significant excess over a 
number of pixels, it is likely that they are real. A detailed discussion 
of the effects of HST breathing and a number of other factors on 
the results of PSF subtraction in the field of quasar LBQS 1210+1731 
is given in section 6 of Kulkarni et al. (2000). 

Object O1 is $\approx 0.2$'' long, while object O2 is 
$\approx 0.15$'' long (just barely resolved), and more diffuse than O1. 
If O1 and O2 are associated 
with the DLA at $z=1.859$, then they are $\approx$ 1.2 and 0.9 
h$_{70}^{-1}$ kpc long, respectively for $q_{0} =0.5$, 
or 1.6 and 1.2 h$_{70}^{-1}$ kpc long for $q_{0} =0.1$. 
No other big, bright galaxies are seen near the quasar. We thus believe 
that the DLA absorber is either compact (e.g. a dwarf galaxy) or a low-surface 
brightness galaxy.

The photometry of O1 and O2 is difficult because of their faintness and 
diffuse nature. We estimated the fluxes by subtracting the PSF star 
from objects O1 and O2, now multiplying the star by factors large enough 
to make objects O1 and O2 look indistinguishable from noise. These PSF 
multiplying factors can then be used directly to estimate the fluxes  
of O1 and O2, since the PSF star P330E is also a well-calibrated NICMOS 
photometric standard. This implies a flux of 1.00 ADU s$^{-1}$ or 
2.19 $\mu$Jy in the F160W filter, before aperture correction. 
To convert the count rate to flux, 
we used the NICMOS photometric calibration factor of $2.190 \times 10^{-6}$ 
Jy/(ADU s$^{-1}$) for the F160W filter, derived 
using the solar-type photometric standard star P330E. 
For object O2, we similarly deduce a flux of 2.85 $\mu$Jy 
in the F160W filter before aperture correction.

Since we had used an aperture of 7.5 pixels to do photometry of P330E, 
we corrected the above flux values of O1 and O2 slightly. 
The aperture correction factor from the flux within a 7.5-pixel radius 
aperture to the total flux has been estimated to be 1.152 for camera 2 filter 
F160W, based on standard NICMOS photometric calibrations made with 
P330E. Using this, we estimate fluxes of 2.52 $\mu$Jy for O1 and 
3.29 $\mu$Jy for O2. For 
reference, the 1 $\sigma$ noise level in the PSF-subtracted image is 
about 0.056 $\mu$Jy per pixel 
in a circular annulus 0.2 $\arcsec$ wide centered at 0.3 $\arcsec$ 
from the quasar center. The corresponding noise levels at 
0.5 $\arcsec$, 0.7 $\arcsec$, 0.9 $\arcsec$, and 
1.1 $\arcsec$ from the quasar center are 0.011, 0.0074, 0.0070, and 
0.0066 $\mu$Jy per pixel respectively. Thus the formal 1 $\sigma$ noise 
uncertainty in the total summed F160W flux over 
the regions occupied by O1 and O2 is $\approx $ 0.11 $\mu$Jy 
(using the noise estimates just outside O1 and O2 at $r=0.3$ \arcsec). 
The errors in the flux values are likely to be larger than this  
estimate since O1 and O2 are barely resolved and are even  
closer to the quasar center. 

The estimated fluxes of 2.52 and 3.29 $\mu$Jy correspond to 
$m_{F160W} = 21.58$ and 21.30, respectively for O1 and O2, 
taking the zero magnitude  
to correspond to 1083 Jy in the Johnson system. 
These observed F160W fluxes correspond to luminosities (at mean rest 
frame wavelength of 5600 {\AA}) of $4.4 \times 10^{9}$ 
h$_{70}^{-2}$ L$_{\odot}$ and $5.9 \times 10^{9}$ h$_{70}^{-2}$ 
L$_{\odot}$ 
respectively for O1 and O2, for $q_{0} =0.5$. For q$_{0} = 0.1$, 
these correspond to  
$8.2 \times 10^{9}$ h$_{70}^{-2}$ L$_{\odot}$ and 
$1.1 \times 10^{10}$ h$_{70}^{-2}$ L$_{\odot}$, respectively. 
Thus, objects O1 and O2 are fainter than an L$_{*}$ galaxy at $z=1.86$ by 
1.9-2.2 magnitudes and 1.2-1.5 magnitudes, for $q_{0} =0.5$ and 
q$_{0} = 0.1$, respectively. If O1 and O2 are not the DLA, the DLA must  
be even fainter.

\subsection{NONCORONAGRAPHIC F187N IMAGES}

At a redshift of $z_{DLA}=1.859$, any H-$\alpha$ emission would be 
expected to lie at $\lambda_{obs}=1.876$ $\mu$m, which is very close 
to the center wavelength $\lambda$ of 1.874 $\mu$m for the filter 
F187N. Thus, the narrow-band images in filter F187N are expected to 
reveal any redshifted H-$\alpha$ emission from the DLA. Fig. 5b 
shows the PSF-subtracted F187N image using the PSF image of the star 
P330E observed on 1997 August 5. The residual image shows a noisy 
feature in roughly the same place and with roughly the 
same size as the feature O1 seen in the 
noncoronagraphic F160W image. But this feature is very weak and 
could be a PSF artifact. No corresponding feature is seen 
for object O2. 

As in the case of the broad-band images, the photometry of O1 
is rather difficult.  Subtracting the 
standard star P330E from O1, scaling the star such that O1 just 
disappears, we estimate a flux of 2.62 $\mu$Jy for O1.  
Here we have used the NICMOS photometric 
calibration factor of $4.107 \times 10^{-5}$ Jy/(ADU s$^{-1}$) 
for the F187N filter. For comparison, 
the 1$\sigma$ noise levels in the F187N image (after PSF subtraction) 
at $r=0.5$ $\arcsec$, 0.7 $\arcsec$, 0.9 $\arcsec$, and 1.1 $\arcsec$ 
from the quasar are 0.033, 0.033, 0.031, and 0.030 $\mu$Jy 
per pixel, respectively. Thus the formal 1 $\sigma$ sky noise uncertainty 
in the total summed F187N flux over 
the region occupied by O1 and O2 is $\approx $ 0.13 $\mu$Jy 
(using the noise estimates just outside O1 and O2 at $r=0.3$ \arcsec). 
Again, the uncertainty in the fluxes is likely to be greater than 
this estimate because O1 is very faint,almost unresolved, and even 
closer to the quasar center. 

The expected F187N continuum must be subtracted from the observed 
flux in order to determine if a statistically significant redshifted 
H-$\alpha$ excess exists. We estimate the continuum under the F187N 
filter by scaling the F160W image using the relative photometric 
calibration of the two filters. 
We find that, in fact, this expected continuum flux agrees almost 
completely with the observed F187N flux. The 1 $\sigma$ noise level in the 
F187N-F160W image is 0.018 $\mu$Jy per pixel just outside the 
location of O1 and O2. This noise level corresponds to a 1 $\sigma$ 
uncertainty of 0.074 $\mu$Jy in the total flux summed over the region 
occupied by O1. We therefore conclude that the contribution to the 
F187N flux from redshifted H-$\alpha$ emission is negligible even for O1. 
It is not likely that we could have missed the H-$\alpha$ 
emission from O1. The H-$\alpha$ emission from the DLA could lie 
outside the F187N bandpass only if the DLA galaxy is lower in 
velocity by more than 1730 km s$^{-1}$ or higher in velocity by 
more than 1280 km s$^{-1}$ 
from the absorption redshift. Such offsets are higher than the 
observed internal velocity dispersion in any typical single galaxy. 

In a 4-pixel region (roughly the size of our resolution 
element), an H-$\alpha$ emission strength of about 0.223 $\mu$Jy 
would yield S/N = 3. With an aperture 
correction factor of 3.37, this corresponds to a total 3 $\sigma$ 
flux limit of 0.75 $\mu$Jy. Integrating over the FWHM of the F187N filter, 
assuming no dust extinction, and using the prescription of Kennicutt 
(1983) for conversion of H-$\alpha$ luminosity to SFR, we get   
a 3 $\sigma$ upper limit on the SFR of 1.3 $h_{0.7}^{-2}$ 
M$_{\odot}$ yr$^{-1}$ for 
$q_{0} =0.5$ or 2.4 $h_{0.7}^{-2}$ M$_{\odot}$ yr$^{-1}$ for 
$q_{0} =0.1$. 

Thus we conclude that the broad-band images suggest possible detections 
of objects O1 and O2 at 0.16'' and 0.24'' from 
the quasar center, although no significant H-$\alpha$ emission is detected 
from either of them. We cannot completely rule out that 
these features could be artifacts of the PSF. In that case, our images 
put very sensitive upper limits on the size 
and brightness of both the DLA absorber and the quasar 
host. We discuss these constraints in section 6. 

\subsection{CORONAGRAPHIC F160W IMAGES}

An F160W coronagraphic image of the central $\sim 3$'' region 
near the quasar is shown in Fig. 6a, 
in which the coronagraphic hole is masked out. Almost 
all the flux seen in this reduced coronagraphic image is due 
to residual  
scattered light from the quasar, and ``glints'' from the 
edge of the hole. After subtraction of a reference PSF image 
using observations of the star GL577, these artifacts 
disappear almost entirely (Fig. 6b).  Features O1 and O2 
seen in Figs. 4b and 5b, are just inside the coronagraphic hole 
and are therefore not seen in Fig. 6b. However, the coronagraph 
is very effective in reducing the quasar light outside of the 
coronagraphic hole, and can therefore be used to look at other 
objects in the field. 

The bright glint immediately to lower left of the edge of the 
coronagraphic hole is unlikely to be real. This is because this 
particular positive ``glint'' has been seen in many other PSF 
subtractions taken at slightly 
different subtractions of other targets where the breathing 
phase differential is slightly negative in target-psf 
subtractions. Furthermore, as noted above, there is a mismatch 
of about 0.5 pixel  between the positions of the PSF star and the 
quasar inside the coronagraphic hole. 
Unfortunately, a better matching PSF star was not available. 
In any case, no other significant objects are seen close to the 
quasar. A number of galaxies are seen farther from the quasar in 
the coronagraphic image (Fig. 3). We will discuss these further 
in section 6.1 below.

\section{DISCUSSION}

The most important result from our observations is that no large bright 
galaxies are seen close to the quasar 
in the field of the DLA absorber toward QSO 1244+3443. Features O1 and 
O2 are the only detected candidates within several arcseconds of the 
quasar and, if real, may be associated with the DLA. 
Their redshift is not confirmed since no H-$\alpha$ emission is detected 
from them. In sections 6.1 and 6.2, we assume that object O1 is associated 
with the DLA to derive constraints on various properties of DLAs. But 
we also consider alternative possibilities in section 6.3, mainly 
the possibility that O1 and O2 may be associated with the host 
galaxy of the quasar. If O1 and O2 are PSF artifacts, then the 
constraints on the DLA and the quasar host are even more severe.

\subsection{CONSTRAINTS ON SIZES, MORPHOLOGY, AND EVNIRONMENTS OF DLAs}

Our observations show no evidence for a big, well-formed galaxy as 
expected in some scenarios for the DLAs [e.g., the proto-spiral model 
suggested by Wolfe et al. (1986), Prochaska \& Wolfe (1997, 1998), 
Jedamzik \& Prochaska (1997)]. Features O1 and O2 have estimated sizes 
of 1-2 $h_{70}^{-1}$ kpc, if they are real and are at the redshift of the 
DLA. This suggests that the absorber is compact 
and clumpy, as expected in the hierarchical picture of DLAs. But it is 
possible that O1 
and O2 are the brightest regions within 
a bigger galaxy, the rest of which we cannot see. Thus, 
we cannot completely rule out the large disk scenario, although 
the compact sizes and low SFRs suggest that the hierarchical picture 
may be favored. Further 
deeper observations will help to more definitively distinguish between 
the large disk vs. hierarchical models. 

Apart from features O1 and O2 very close to the quasar, 
our images show several galaxies in the  
F160W and F187N images. In particular, the coronagraphic F160W image 
shows 7 galaxies, 1 of which is very bright while 3 are moderately bright 
galaxies aligned in almost a 
straight line with the quasar, roughly to the east of the quasar. The other 
three are fainter. The closest galaxy in roughly the east direction  
is also seen in the noncoronagraphic F160W image, and barely seen in the F187N 
image. The other galaxies seen in the coronagraphic F160W image are outside the 
field of the non-coronagraphic images and thus not seen in them. 
From the galaxy number count-magnitude relation based on deep NICMOS images 
(Yan et al. 1998), about 1 galaxy is expected for $H < 21$ in the camera 2 field. 
In comparison, the high density of galaxies in our NICMOS images suggests that 
several of these galaxies may belong to 
the same group as the DLA, and clustering may be enhanced near this DLA. (However, 
we do not have redshift information on these galaxies.) In any case, they have 
fairly large angular separations from the quasar (3.9 $\arcsec$, 6.7 $\arcsec$, 9.2 $\arcsec$, 11.3 $\arcsec$, 12.0 $\arcsec$, 14.6 $\arcsec$, and finally 
16.8 $\arcsec$ for the bright galaxy) suggesting it is unlikely for any of 
them to be the DLA absorber itself. 
 
\subsection{CONSTRAINTS ON STAR-FORMATION RATE AND DUST IN DLAs}

The lack of significant rest-frame H-$\alpha$ emission in our 
images puts fairly tight constraints on the star-formation rate 
in the DLA toward QSO 1244+3443, i.e. a $3 \sigma$ upper limit of 
1.3 (2.4)  h$_{70}^{-2}$ M${_\odot}$ yr$^{-1}$ for 
$q_{0} = 0.5 (0.1)$, if no dust is assumed. 
This is similar to (even smaller than) the upper limit on the SFR in 
the high-$z$ DLA toward LBQS 1210+1731 (Kulkarni et al. 2000), i.e., 
a 3 $\sigma$ constraint of 4 h$_{70}^{-2}$ M${_\odot}$ yr$^{-1}$ for 
$q_{0} = 0.5$ or 7.4 h$_{70}^{-2}$ M${_\odot}$ yr$^{-1}$ for 
$q_{0} = 0.1$. For comparison, the near-IR spectroscopic survey of 
Bunker et al. (1999), aimed at detecting H$\alpha$ from DLAs, gave  
typical upper limits of $\approx$ 15 M$_{\odot}$ yr$^{-1}$, for 
q$_{0}=0.5$ and $H_{0} = 70$ km s$^{-1}$ Mpc$^{-1}$. 
 Our limits on the SFR in QSO 1244+3443 and LBQS 1210+1731 mark 
a big improvement over the tightest previous constraints on the SFR 
in DLA galaxies from H-$\alpha$ spectroscopy. (See Fig. 19 of Kulkarni 
et al. 2000 for a  detailed comparison.)

In principle, the lack of detectable H-$\alpha$ emission from the 
DLA could be because of dust extinction, in which case 
the actual SFR could be higher. However, if the HI-column 
density weighted average SFR was that predicted by the closed box model of 
Pei \& Fall (1995) and if DLAs are large disks, with the global SFR 
distributed among them, then to reconcile the prediction  
of 38.5 M$_{\odot}$ yr$^{-1}$ at $z=1.86$ to a value below our upper limit, 
one would require an optical depth $\tau_{0.66 \mu m} > 3.4$. (See Fig. 19 
of Kulkarni et al. 2000 for the predictions of the closed-box 
model for large disks.) This implies an optical depth at 4400 {\AA} of 
$\tau_{B} > 5.8$ and hence 
a mean dust-to-gas ratio $k \equiv \tau_{B} (10^{21}/N_{HI}) > 14.6$. 
Here we have assumed an extinction curve similar to that of the Milky Way or 
the Small Magellanic Cloud or the Large Magellanic Cloud. 
Dust-to-gas ratios $k > 14.6$ are much higher  
than the mean dust-to-gas ratio of 0.8 for the Milky Way, 
or the typical value of $\sim$ 0.03-0.1 for the DLA galaxies, suggested by 
observations of background quasar reddening and heavy element depletions 
(see, e.g., Pei et al. 1991; Pettini et al. 1997 and references therein). 

One could ask whether the DLA may be simply 
hiding because parts of it could be very dusty. This is a possibility since 
 the regions of DLAs probed by spectroscopy of quasars may be 
systematically less dusty. Such a selection effect could arise 
because the dustier regions would extinguish 
the quasar (see, e.g., Fall \& Pei 1993). However, 
as discussed above, it would take a dust to gas ratio 
of greater than 14.6 (and $A_{H-\alpha} > 3.7$ magnitudes)  to extinguish 
the H-alpha emission expected at the average SFR predicted for a 
large-disk DLA at z=1.86.  Thus if the regions of the DLA away from  
the quasar sightline were to have their H-alpha emission extinguished 
by dust, the dust-to-gas ratio would have to change by a factor of 
several hundred from the line-of-sight to the quasar to all 
off-line directions. Such a situation is possible, but seems rather 
contrived. In this context, we note that Glazebrook 
et al. (1999) observed a sample of 13 Canada-France 
Redshift Survey galaxies at $z \sim 1$ in redshifted H-$\alpha$. 
They found that the SFR from H-$\alpha$ is $\sim 3$ times that inferred 
from the UV and that the extinction is moderate, with $A_{V} \sim 1$ at 
most for these galaxies. Kennicutt (1998) discusses the merits of and 
the systematic errors in using H-alpha to derive SFRs. He reports a mean 
extinction $A_{H-\alpha}$ of 0.5-1.8 magnitudes for large samples of H II 
regions in nearby galaxies. Thus there is no reason to expect the $z=2$ 
DLAs to have extinction much higher than these amounts. It seems more 
natural to interpret the lack of H-$\alpha$ 
in terms of a low star formation rate assuming reasonable numbers for 
the dust-to-gas ratio. The real picture may be a combination of low 
star formation rate and some (moderate) amount of dust extinction. 
If the SFR is indeed low, then this together with 
the compact sizes seen in our images suggests 
that the DLA may be a dwarf galaxy or a low 
surface brightness (LSB) galaxy, rather than a large bright disk. 

To compare our relative sensitivity to detection of various kinds of 
galaxy morphologies, we have carried out simulations for disk and LSB 
galaxies.  Using our observed noncoronagraphic F160W image of 
QSO 1244+3443, we created a 
simulated image by adding disks or LSBs of varying brightnesses 
at angular separations of 
1.0 $\arcsec$ from the quasar (taken as a representative angular separation 
of interest). The simulated images were made using the IRAF task MKOBJECTS 
and then PSF-subtracted in a manner similar to 
that used for the actual quasar image. The magnitudes of the galaxies were 
varied to see how faint these objects have to be in order not to be 
detected in our F160W NICMOS image even after PSF subtraction. 
A disk with an exponential surface brightness profile and a scale 
length of  3 $h_{70}^{-1}$ kpc was placed at
a distance of 1" in projection from the quasar.  Such a disk galaxy
would be barely detectable in our observations if the disk had H=22,
and more easily detectable with H $ < 21$. For a low surface brightness 
galaxy, 
we assume an exponential brightness profile with a scale length of 
14  $h_{70}^{-1}$ kpc (i.e. 10  $h_{100}^{-1}$ kpc, the average scale 
length of a giant LSB galaxy --see Sprayberry et al. 1995). We find 
that such a giant LSB galaxy when placed 1 $\arcsec$ 
away from the quasar would be hard to detect in our F160W image even after 
PSF subtraction, if it had an integrated H magnitude of $H \ge 19$. On 
the other hand LSBs with H $\le 18.5$ would be detectable at separations 
of 1 $\arcsec$ from the quasar (more easily so for H $\le 18$). 

We also note that an LSB galaxy with H = 18 separated 1 $\arcsec$ away 
from the 
quasar would have been easy to detect and would in fact look roughly similar 
(in shape, size and brightness) to the halo left behind around the quasar 
after PSF subtraction in our actual observed image (Fig. 4b). Thus it would 
be hard to distinguish between the 
PSF subtraction residuals and an LSB of H=18 positioned exactly on top 
of the quasar. However, since the circum-quasar residuals seen in our 
F160W image correspond well in positions of detailed features with 
the features in the PSF of a star like P330E, we take the view that the 
faint halo seen around the quasar after PSF subtraction (Fig. 4b) is indeed 
because of PSF subtraction residuals rather than an LSB  
galaxy of H = 18.  This and the fact that no H $ \ge 19$ LSBs are 
detectable clearly underscore the possibility that our NICMOS F160W image 
could be hiding a large LSB at angular separations as close as (or closer 
than) 1 $\arcsec$ from the quasar. 

We expect that in comparison to LSBs, compact 
objects should be much easier to detect. At the close 
angular separations of features O1 and O2 from the quasar, we expect the 
detection of disks and LSBs to get harder than estimated in the above 
simulations. Features O1 and O2 have H = 21.58 and 21.30 respectively, 
somewhat brighter than the H=22 limit of detection for a normal 
disk 1 $\arcsec$ away, but much fainter than the detection limit 
of H = 19 for an LSB 1 $\arcsec$ away. Thus, if these features are real, 
they may be compact, dwarf galaxies themselves or 
they could be parts of larger disks or LSBs. The lack of any other objects 
in regions further out (at separations comparable to 1 $\arcsec$ from the 
quasar) can rule out presence of disks with $ H \le 22$ or LSBs with 
$ H \le 19$. The limits are even tighter for separations out to 
6.7 $\arcsec$ which is the separation of the nearest galaxy detected in 
the observed non-coronagrahic F160W image, or even out to 3.9 $\arcsec$, 
the separation of the nearest galaxy detected in the coronagraphic F160W 
image.

\subsection{ALTERNATIVE POSSIBILITIES}

Finally, if O1 and O2 are real, it is possible that they arise 
in the quasar host galaxy, rather than the DLA. We cannot test this 
possibility further because we do not have narrow-band images in 
filters tuned to $z_{em}=2.48$. 
However, we cannot rule out this possibility either. If O1 and O2 are in 
fact associated with the host galaxy of the quasar, then they would have 
luminosities (at rest frame 0.46 $\mu$m) 
of $\approx 8.4 \times 10^{9}$ $h_{70}^{-2}$ L$_{\odot}$ and 
$1.1 \times 10^{10}$ $h_{70}^{-2}$ L$_{\odot}$ respectively, 
for $q_{0} =0.5$ ($1.8 \times 10^{10}$ $h_{70}^{-2}$ L$_{\odot}$ 
and $2.4 \times 10^{10}$ $h_{70}^{-2}$ L$_{\odot}$, respectively,  
for $q_{0} =0.1$).
The images would then suggest that the quasar host is not a 
large galaxy with or without interactions, but rather shows a compact 
morphology. The strongest feature in the quasar host would then be 
off-center with respect to the quasar nucleus, which has been observed 
in other quasars. If O1 and O2 are  
in fact the quasar host, then the limits on the luminosity and SFR in 
the DLA are even more severe than our estimates in sections 5.1 and 
5.2. Conversely, if the features are associated with the DLA galaxy, 
then the constraints on the 
quasar host are more severe than those given above. 

It is also possible that we may be seeing interloper 
galaxies at redshifts other than that of the DLA. In this context, we note 
that ground-based spectra of QSO 1244+3443 have revealed 3 other absorbers 
besides the DLA, at $z =$ 1.8444, 1.8491, and 1.9126 (Wolfe et al. 1993). These absorbers 
may explain some of the galaxies seen at larger angular separations from 
the quasar (Fig. 3). Indeed the absorbers at $z = 1.8444$ and 1.8491 may 
belong to the same group of galaxies as the DLA. Our narrow-band searches, 
which failed to detect any line emission at all, would have been able to 
detect rest-frame H-$\alpha$ emission 
from even the absorbers at $z = 1.8444$ and 1.8491. We do not have H-$\alpha$ 
images at the redshift of the absorber at $z = 1.9126$. Thus 
we cannot test the interloper possibility further. However, since  
the other absorbers are not DLAs and have weaker Ly-$\alpha$ and metal 
absorption lines than the DLA, they seem less likely to be detectable 
than the DLA itself. 

\subsection{COMPARISON WITH OTHER WORK}

Recently, Warren et al. (2000) have presented analysis of a NICMOS 
broadband imaging survey in the F160W filter for 15 quasars with DLAs. 
Their method of PSF subtraction is different from ours. For each quasar 
in their sample, they use the average of their other 14 observed quasar 
images to create a PSF for subtraction. This has the advantage of achieving 
a good color match (as compared to using a stellar PSF) and also having the 
PSF image land upon nearly the same subpixel locations as for the 
quasar. However it does 
reduce the possibility of detecting the quasar host galaxies or diffuse 
(e.g low surface brightness) galaxies at very small angular separations 
from the quasar. We take the approach of using the observed PSFs of stars, 
taking care of pixel nonlinearization effects in data reduction with 
NICRED and subpixel positioning accuracy between quasar and PSF star in 
IDP-3. This allows us to be more sensitive to detecting quasar hosts or 
diffuse objects very close to the quasar. Warren et al. estimate that 
the accuracies of the two PSF subtraction techniques are comparable 
(probably more so   
at larger angular separations--above $\approx 0.5-0.6$''.) Indeed features  
O1 and O2 here and in LBQS 1210+1731 (Kulkarni et al. 2000) appear comparable 
to the object N12-1D of Warren et al. (2000). We believe that object could 
also be a DLA, either a compact object or the brightest part of a bigger 
galaxy; alternatively, it could also be the quasar host galaxy. Overall, 
our results are consistent with those of 
Warren et al., who also find no large bright galaxies in their NICMOS survey. 

\section{CONCLUSIONS AND FUTURE WORK}

Our continuum and H$\alpha$ images of the $z=1.86$ DLA toward 
QSO 1244+3443 suggest that this DLA is not a big 
galaxy with high SFR, but may be compact (1-2 $h_{70}^{-1}$ kpc in 
size), probably consisting of multiple sub-units. A few possibly companion 
galaxies are seen at larger angular separations in the continuum  
images. Assuming no dust 
extinction of H-$\alpha$ emission, we place a 3 $\sigma$  upper limit of 
1.3 (2.4) $h_{70}^{-2}$ M$_{\odot}$ yr$^{-1}$ on the star formation rate, 
for $q_{0}=0.5 (0.1)$. Our observations 
are consistent with the hierarchical models, in which DLAs arise in 
several sub-galactic clumps or dwarf galaxies, which eventually come 
together to form the present-day galaxies (see, e.g., York et al. 1986;  
Matteucci et al. 1997). Indeed, theoretical simulations of 
merging proto-galactic fragments in cold dark matter cosmologies  
(e.g., Haehnelt et al. 1998) and collapsing halos with merging clouds  
(e.g., McDonald \& Miralda-Escud'e 1998) have been found to reproduce 
the observed asymmetric line profiles in metal absorption lines of DLA 
galaxies.  However, it cannot be ruled out that 
the DLA toward QSO 1244+3443 is a large low-surface brightness galaxy 
with a low SFR, the rest of which is below our detection limit even in 
the F160W image. 

Our results for QSO 1244+3443 agree with our earlier results for LBQS 
1210+1731 and results for our remaining NICMOS observations 
(Kulkarni et al. 2000, 2001). The findings of the NICMOS broad band 
imaging study by Warren et al. (2000) also agree with our broad-band 
results here. In the end, to verify the reality of all our candidate 
objects and similarly those of Warren et al., confirming spectroscopy 
or narrow-band imaging is necessary. The lack of significant H-$\alpha$ 
detections in our observations suggests that it will be very helpful 
to have deep Ly-$\alpha$ imaging with STIS, or deeper near-IR narrow-band   
imaging with NICMOS when it becomes available again. 
Narrow-band imaging observations will help to establish 
the redshift identifications of the DLAs, their companion galaxies, 
and quasar host galaxies, and are likely  to be more efficient than 
spectroscopy. It is also necessary to increase the number of imaging studies  
of high-$z$ DLAs, since the current sample of DLAs studied 
at high resolution to search for H-$\alpha$ emission 
is still small. It is quite possible that 
different DLAs have different rates of evolution because of different 
physical conditions. Indeed, this is suggested by the large scatter in the 
metallicity-redshift relation of 
DLAs (see, e.g., Pettini et al. 1998 and references therein). 
To improve the statistics of the DLA imaging studies, it is necessary 
to obtain high spatial resolution near-IR images of 
more high-redshift DLAs. A major advantage of future HST observations 
will be relatively stable PSFs compared to those currently 
achieved with ground-based 
telescopes, a factor that is crucial for the detection of DLAs. 
It will also be of great interest to complement the HST observations 
with observations from adaptive optics systems on 
large ground-based telescopes. Although these systems do not currently 
have the 
relatively stable PSF offered by HST, they will be able to achieve 
even higher spatial resolution and higher imaging sensitivity. Such 
combined future space and ground-based observations will provide further 
insight into the structure and nature of DLAs and their relation 
to other galaxies.  

\acknowledgments

This project was supported by NASA grant NAG 5-3042 to the NICMOS 
Instrument Definition Team. It is a pleasure to thank Nicholas Bernstein 
and Keith Noll for their assistance in the scheduling of 
our observations. We thank Elizabeth Stobie, Dyer Lytle, Earl O'Neil, 
Irene Barg, and Anthony Ferro for software and computer support. 

\clearpage

\clearpage

\centerline{\bf FIGURE CAPTIONS}

{\bf FIG. 1--} NICMOS camera 2 noncoronagraphic 1.6 $\mu$m broad-band 
image of the field of QSO 1244+3443. The color scheme is indicated 
with the bar on the bottom of the image. 
Image Y axis is -136.943 degrees east of north.

{\bf FIG. 2--} NICMOS camera 2 noncoronagraphic 
1.9 $\mu$m narrow-band image of the field of 
QSO 1244+3443. Image Y axis is -136.943 degrees east 
of north.

{\bf FIG. 3--} NICMOS camera 2 coronagraphic 1.6 $\mu$m broad-band 
image of the field of QSO 1244+3443. 
The quasar has been placed in the coronagraphic hole.
Image Y axis is -117.897 degrees east of north.

{\bf FIG. 4--} Zoomed-in $2.81 $\arcsec$  \times 2.79 $\arcsec$ $ region 
of the NICMOS camera 2 noncoronagraphic 1.6 $\mu$m broad-band 
image of the field of QSO 1244+3443, (a) before PSF subtraction (top), 
(b) after PSF subtraction (bottom).

{\bf FIG. 5--} Zoomed-in $ 2.81 $\arcsec$  \times 2.79 $\arcsec$ $ region 
of the NICMOS camera 2 noncoronagraphic 1.9 $\mu$m narrow-band image of 
the field of QSO 1244+3443, (a) before PSF subtraction (top), (b) after 
PSF subtraction (bottom).

{\bf FIG. 6--} Zoomed-in $2.81 $\arcsec$  \times 2.79 $\arcsec$ $ region 
of the NICMOS camera 2 coronagraphic 1.6 $\mu$m broad-band image of the 
field of QSO 1244+3443, (a) before PSF subtraction (top), (b) after 
PSF subtraction (bottom).

\end{document}